\documentclass[12pt, a4paper]{article}
\pdfoutput=1
\usepackage{blindtext}
\usepackage{color}
\usepackage{amssymb}
\usepackage{titlesec}
\usepackage{graphicx}
\usepackage{dcolumn}
\usepackage{float}
\usepackage{appendix}
\usepackage{subcaption}
\usepackage{graphicx}
\usepackage{mathrsfs}
\usepackage{tikz}
\usepackage{caption}
\usepackage{siunitx}
\usepackage{listings}
\usepackage{xcolor}
\usepackage{lstautogobble}
\usepackage{hyperref}
\usepackage{authblk}
\usepackage{bm}
\usepackage[margin=1.0in]{geometry}

\definecolor{codegray}{rgb}{0.5,0.5,0.5}

\hypersetup{
	colorlinks=true,
	linkcolor=black,
	urlcolor=blue
}

\lstMakeShortInline!

\begin{document}
	\title{The Einstein Toolkit: A Student's Guide}
	\author{Nicholas Choustikov\thanks{nc538@cam.ac.uk}}
	\affil{Institute of Astronomy, University of Cambridge, Cambridge, CB3 0HA, United Kingdom}
	\date{October 21, 2020}
	\maketitle
	
	\hrule
	\begin{abstract}
		The Einstein Toolkit represents a unique opportunity for students to explore the world of numerical relativity, without the need for high-level computing power or knowledge of the mathematics behind the simulations themselves. This document aims to act as a resource for students aiming to learn about and get involved with Numerical Relativity, focusing specifically on the modelling of Binary Black Hole Mergers in order to produce sample Gravitational Waveforms as would be measured by Gravitational Wave detectors such as LIGO.
		
		The feasibility of using the Einstein Toolkit on Amazon's Web Services (AWS) has also been explored for the first time, confirming a potential - yet expensive - method for students or citizen scientists to gain access to high-level computing power sufficient to play with and run their own numerical relativity simulations.
	\end{abstract}
	\hrule
	\vspace*{1 cm}
	
	\tableofcontents
	\newpage
	
	\section{Introduction}
	
	The Einstein Toolkit (ET) is a community-driven platform of core computational tools \cite{Loffler_2012} which provide a powerful tool to perform high-level numerical-relativity and relativistic astrophysical simulations, even while operating within the constraints of desktop-level computational power \cite{Loffler_2012, ZILH_O_2013}. Such a toolkit appeals therefore as a particularly useful tool for students beginning to learn and use Numerical Relativity (NR) within the context of university-level physics, as well as to researches carrying out advanced simulations. 
	
	Therefore, this tutorial has been written with a keen undergraduate student in mind. The intention being to set up said student with the ability to utilize the toolkit, as well as to understand what goes into parameter files used for simulations - whether for a research project or out of personal curiosity. As a result, this tutorial will \textit{not} require a strong understanding of General Relativity (GR), though a basic understanding of the astrophysical phenomena being studied is incredibly useful.
	
	This tutorial will first go through the entire process of installing, building and then using the Einstein Toolkit - assuming little knowledge of shell scripting - or of the toolkit itself. In section \ref{6} it will then go through a sample parameter file (.rpar), allowing the reader an opportunity to become familiar with parameter files as well as the Cactus Thorn structure \cite{ZILH_O_2013}. The process of accessing and parsing data produced following a successful run is then discussed in section \ref{data}. For more experienced readers, it is possible to follow the Einstein Toolkit's own tutorial, available with an executable jupyter notebook available at \url{https://einsteintoolkit.org/documentation/new-user-tutorial.html}.
	
	\section{Requirements and Feasibility}
		
	Naturally, the equations being solved in NR are computationally expensive \cite{Expensive}. As a result, despite the fact that the ET has been well-optimized for even desktop-level computational power, there are certain computational requirements.
	
	In principle, the Toolkit can be built and run on systems with one processor and 4GB of RAM. As will be discussed, often high-level, accurate simulations can easily become computationally expensive to the point of effective impossibility on such systems, though alternatives are suggested.
	
	Furthermore, in order to ease installation and use of the toolkit, it is recommended to use a Linux distribution. All examples in this tutorial were performed using a clean system running the minimal Ubuntu 20.04 distribution \cite{Ubuntu}. For such a clean system, the following packages are required (with their most-recent versions) in order to build and run the ET, with a sample of the required bash code shown below:
	
	\begin{center}
		\begin{tabular}{|| c | l ||}
			\hline
			Package	&	Installation Code	\\
			\hline\hline
			Git	&	!sudo apt install git!	\\
			\hline
			Subversion	&	!sudo apt-get install subversion!	\\
			\hline
			Curl	&	!sudo apt install curl!	\\
			\hline
			Python2/3	&	!sudo apt install python2/3!	\\
			\hline
			GCC	&	!sudo apt-get install build-essential!	\\
			\hline
			GFortran	&	!sudo apt-get install gfortran!	\\
			\hline
			MPI	&	!sudo apt-get install mpi!	\\
			\hline
			
		\end{tabular}
	\end{center}

	Next is a question of feasibility. As mentioned before, many high-level simulations will require literally thousands of core-hours, and as a result can quickly grow to become too computationally-expensive to run feasibly on a personal-computer. For instance, using the models that will be discussed in sections \ref{Examples} and \ref{6} as examples, the TOV Star Simulation requires approximately 10 core-minutes, while the Binary Black Hole (BBH) Merger requires 8700 core-hours. This clearly renders it rather expensive to run on a laptop. Therefore, in cases where the user does not have access to a computing cluster, the best option - despite being fiscally expensive - might potentially be to make use of the Amazon Web Services (AWS) cloud-computing service.
	
	Available to members, the AWS Elastic Computing Cloud (EC2) is a service which provides highly-customizable computing capacity in the cloud. This is done, by first choosing the 'shape' of a desired virtual machine (VM). This decides the CPU and RAM size available to the machine, from a list of options pre-determined by AWS. Following this, an image (custom or default) is chosen which will run on the VM. For the purpose of this project, a public image (freely available) running a clean installation of Ubuntu 20.04 \cite{Ubuntu} was chosen, to be run on an m5.16xlarge instance. This setup offered performance equivalent to 64 CPUs with 256 GiB memory \cite{AWS}, at the cost of \$3.072 per hour. This though can quickly become equally prohibitively expensive. In order to combat this, a system was setup with a large detachable volume which could be attached to the primary instance while calculations were being run, and then attached to a much smaller (and cheaper) instance in order to then carry out data analysis.
	
	This option can equally become particularly expensive, but it certainly offers an opportunity for students with external funding to play with computing power previously only available in dedicated computing clusters.  
	
	During the process of writing this document, several models of BBH mergers, with varying mass ratios, initial separations and intrinsic spins were started and run through the first 20,000 iterations. These yielded the expected behaviour and waveforms up to this point in the model. None were however run to completion due to a lack of sufficient funding.
	
	\section{Installing and Building the Toolkit}
	
	Once all necessary modules have been acquired, the following steps should be taken in order to install the ET. It is important to note, that in cases where only one user will be utilising the toolkit, that user itself should run all setup code. If on the other hand multiple users will be interacting with the toolkit, then the root user should not be used. This is specifically important due to the fact that by default, all simulation data will be stored in the user's home directory, as described in section \ref{data}.
	
	To start the process, the following three commands should be called, in the directory within which the toolkit will be installed:
	\begin{lstlisting}
	curl *-*kl0 https://raw.githubusercontent.com/gridaphobe/CRL/ET_2020_05/GetComponents
	chmod a+x GetComponents
	./GetComponents --parallel https://bitbucket.org/einsteintoolkit/manifest/raw/ET_2020_05/einsteintoolkit.th
	\end{lstlisting}
	
	This code downloads the !GetComponents! package, which is then executed in order to install the ET itself. Following this, chmod allows the package to be executed, and the third line executes it, providing the !einsteintoolkit.th! file as an argument, to be used as the thornlist. Once this is done, in principle all component thorns of the ET have been acquired, they just need to be configured for a given machine, such that they will be able to interact with machine-specific queuing systems.
	
	This is principally done using the Simulation Factory ("SimFactory"), which itself is configured as shown: 
	\begin{lstlisting}
	cd Cactus
	./simfactory/bin/sim setup*-*silent 
	\end{lstlisting}
	Once this step has been completed, SimFactory has been configured to run using the machine's default settings, which can be accessed and edited in !./simfactory/mdb/machines/<hostname>.ini!. Any changes here should be carried out with care, as they may lead to warnings or errors when simulations are started-up. Furthermore, it is important to note that by default the toolkit is configured to run with only $1$ node.
	
	Now, the ET must be built. Assuming that there are no issues with the SimFactory setup, and that all required packages are functioning properly, the next command should completely build the toolkit.
	
	\begin{lstlisting}
	./simfactory/bin/sim build *-*j<N> *-**-*thornlist ../einsteintoolkit.th
	\end{lstlisting}
	where $N$ is the number of processors used when building, and should be changed to suit the given machine. It should be noted that even with multiple processors, this step can take a very long time. For example, with $N = 2$, it took approximately three and a half hours to complete, while with $N = 64$, it took fewer than 15 minutes to completely build.
	
	Assuming this step finished with no issues, the toolkit has now been built, and is ready to run several test examples, as described in section \ref{Examples}.
	
	\section{Interacting with the Toolkit}
	
	The toolkit is made up of a series of Cactus components, or thorns, which for all intents and purposes can be thought of as modules which can be called upon to accomplish various tasks. The complete thornlist along with all documentation is available at \url{https://einsteintoolkit.org/documentation/ThornGuide.php}.
	
	Tasks achieved by these thorns include multiple vacuum spacetime solvers, relativistic hydrodynamic solvers, initial data components, along with others for data output and analysis.
	
	Therefore, in order to run a simulation, it is necessary to write a parameter file (.par or .rpar) which tells the toolkit which thorns to use, along with the desired parameters - all working together to produce the required model. As a result, it quickly becomes clear that dependant on the system being investigated, different thorns will prove far more efficient and useful - especially compared to thorns which on the surface appear to accomplish the same task.
	
	Once a parameter file has been written, all interaction with the ET can be done using SimFactory. In principle, the user will instruct SimFactory to create and run a simulation, following a given parameter file. This can be achieved using the following command:
	\begin{lstlisting}
	lamboot
	./simfactory/bin/sim create-run <SIM_NAME> <PARFILE> <N> <WALLTIME>
	\end{lstlisting}
	
	It is important to note that when using a computing cluster, it is necessary to call !lamboot! every time an instance running the ET is started, as otherwise errors will be encountered when the simulation is started.
	
	This code will create a simulation under the name !<SIM_NAME>!, and then proceeds to run it, using !<N>! processors with a run time of !<WALLTIME>!.
	
	More useful, is an equivalent command to create and submit a given simulation:
	\begin{lstlisting}
	./simfactory/bin/sim create-submit <SIM_NAME> <PARFILE> <N> <WALLTIME>
	\end{lstlisting}
	this will add the simulation to the queue, telling SimFactory to begin running it as soon as there are sufficient resources available. It is also particularly useful, as in doing so the simulation will run in the background - therefore allowing the user to do something else in the meantime. Such as:
	\begin{lstlisting}
	./simfactory/bin/sim list-simulations
	./simfactory/bin/sim show-output <SIM_NAME> --follow
	./simfactory/bin/sim stop <SIM_NAME>
	\end{lstlisting}
	as expected, the first command will list all simulations in the toolkit's memory, along with their statuses (!RUNNING/FINISHED!). The next command will show the full terminal output of the simulation, with live output. This can be stopped with !<Ctrl + C>! without stopping the simulation itself. Alternatively, the !--follow! parameter can be removed, in which case the entire output history of the simulation up to that point will be printed. This allows the user to see how the simulation is proceeding. Finally, the third command will terminate the simulation inelegantly, if required.
	
	There are a plethora of other commands available for working with SimFactory, and these can be found and explored on the SimFactory documentation \cite{SimFactory}. 
	
	\section{Example Parameter Files}\label{Examples}
	
	As is historically known, often the most simple test of a running system is to produce a !Hello World! output. For this purpose, there is !HelloWorld.par!, which is found within the toolkit at !~/Cactus/arrangemets/CactusExamples/HelloWorld/par/HelloWorld.par!. On running this "simulation", if the output "Hello World" is seen printed anywhere, it can be considered a successful test - and as a result the toolkit is considered to be running correctly.
	
	Following this, a further - more relevant - example simulation is that of a simple TOV Neutron Star (NS).
	
	Here, the simulation evolves a single, stable, non-rotating NS using a solution to the Tolman-Oppenheimer-Volkoff equation \cite{TOV}	
	\begin{equation}
	\frac{d P}{d r} = - \frac{G m}{r^2} \rho \bigg{(} 1 + \frac{P}{\rho c^2} \bigg{)} \bigg{(} 1 + \frac{4 \pi r^3 P}{m c^2} \bigg{)} \bigg{(} 1 - \frac{2 G m}{r c^2} \bigg{)}^{- 1}
	\end{equation}
	as the initial condition. Furthermore, the model assumes a polytropic equation of state $P = K \rho^{\Gamma}$, with $\Gamma = 2$ and $K = 100$. 
	
\begin{center}
		\begin{figure}[h!]
			\includegraphics[width=0.99\linewidth]{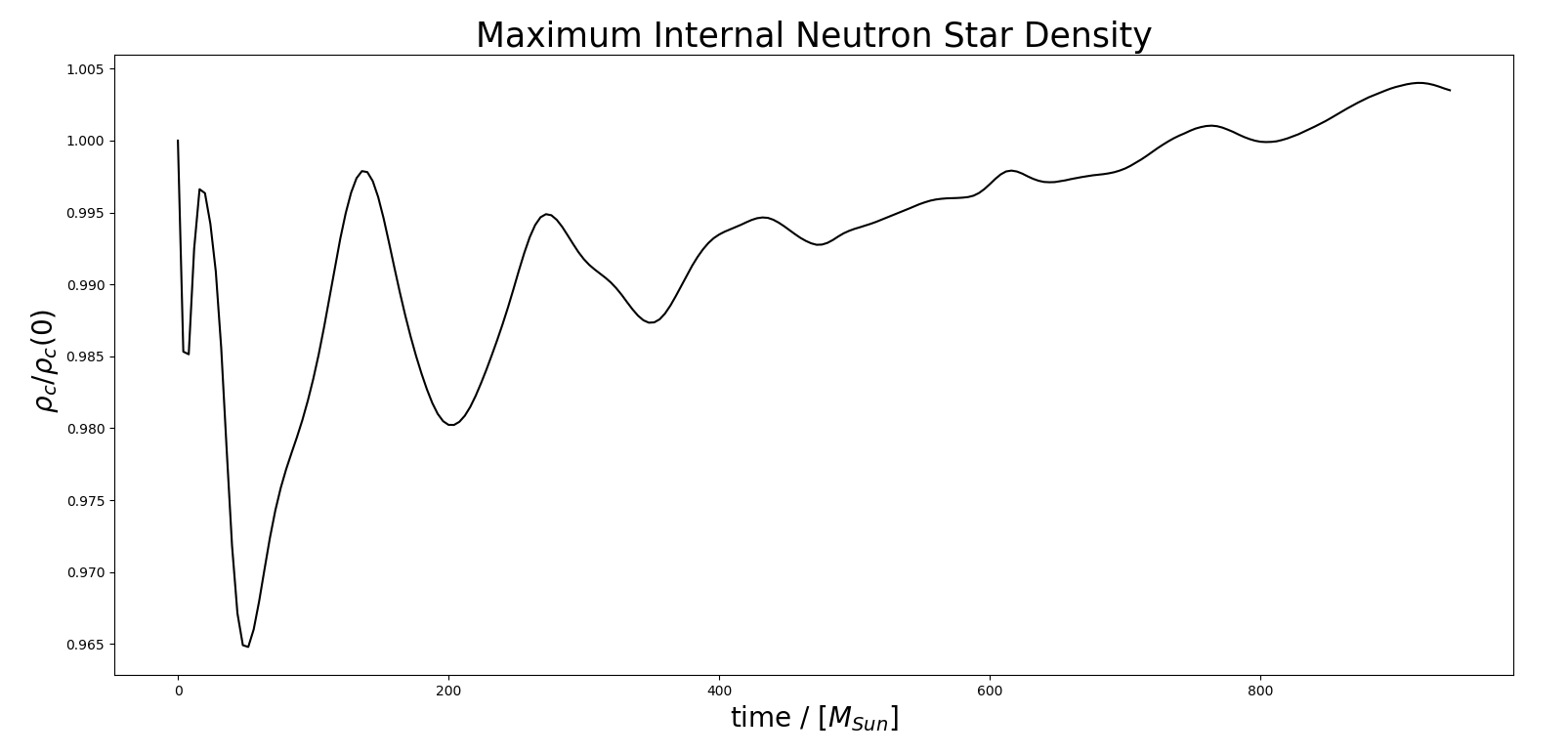}
			\caption{\protect\label{rhomax}Maximum internal density within a non-rotating neutron star of mass $M = 1.4 M_{\odot}$ plotted as a function of time. Model was produced using the Einstein Toolkit, by numerically evolving a system from initial conditions produced as a spherically symmetric solution to the TOV equation for a compact neutron star.}
		\end{figure}
	\end{center}

	Despite the fact that this is a fairly low-resolution model, it serves as a reasonable representation of a non-rotating NS with a mass of $M = 1.4 M_{\odot}$ and as a result a reasonable test for the ET, without being too computationally intensive. The simulation's maximum density ($\rho_c$) has been plotted in figure \ref{rhomax}. Here, the observed oscillations are produced by interactions between the star and the artificial atmosphere \cite{Loffler_2012}.
	
	The parameter file used, along with a further description and documentation is available at \url{https://einsteintoolkit.org/gallery/ns/index.html}.
	
	\section{Diving into a Black Hole Binary} \label{6}
	\subsection{The Merger}
	
	Since the advent of effective numerical relativity, one particularly successful topic of research has been that of BBH mergers. Such systems are typically produced by either dynamical means, whereby dynamical interactions between black holes in dense clusters can lead to binaries forming \cite{Banerjee_2017}, or by non-dynamical means, whereby binaries made up of two massive stars survive as a close binary due to their tidally locked high spin, until they both collapse into black holes \cite{MOB}. 
	
	These systems evolve as follows. First, the two black holes exist as a classical binary, with the two compact bodies orbiting each other. As they orbit, Gravitational Waves (GW) are emitted, acting as an energy-loss function \cite{basics}. This leads to the two bodies orbiting closer to and faster around each other, in the inspiral phase. Eventually, the two collide and merge - producing the GW 'chirp' \cite{LIGOObs} - lasting a fraction of a second - followed by the ringdown phase \cite{Bhagwat_2018}. Measurements of the GWs produced throughout the system's evolution can yield valuable information about the two bodies involved, as well as the exact nature of their orbit. This data can therefore provide an extraordinary opportunity to test GR under such high-energy conditions \cite{EMRI}.
	
	In principle, the inspiral and ringdown can be modelled using Post-Newtonian approximations \cite{unreasonable} and black hole perturbation theory \cite{Le_Tiec_2014} respectively. However in order to fully model the entire evolution of a given system, the full equations of GR must be solved, and as a result NR truly comes into its own.
	
	Obtaining accurate models of BBH mergers has become incredibly important, with the recent construction of gravitational wave observatories such as LIGO \cite{Abbott_2009} and VIRGO \cite{VIRGO} paving the way for several observational runs for which sample waveforms are naturally needed. One of the most important results came in 2015 with the detection of GW150914 \cite{LIGOObs}, the first successful detection of gravitational waves due to a BBH merger, coming a century after Einstein's prediction. This system involved two black holes with initial inferred masses of $36 M_{\odot}$ and $29 M_{\odot}$ at a distance of $\SI{410}{\mega pc}$ from Earth \cite{LIGODATA}. It is a model of this very event which is to be discussed here.
	
	\subsection{The Model}
	
	The simulation discussed in this section can be found at \url{https://einsteintoolkit.org/gallery/bbh/index.html}, with the simulation's parameter file available at \url{https://bitbucket.org/einsteintoolkit/einsteinexamples/raw/master/par/GW150914/GW150914.rpar} \cite{wardell_barry_2016_155394, Loffler:2011ay, Pollney:2009yz, Thomas:2010aa, Brown:2008sb, Husa:2004ip, Thornburg:2003sf, Schnetter:2003rb, Ansorg:2004ds, Goodale:2002a, Dreyer:2002mx}. 
	
	Investigating the parameter file, it is clear to see that the file's preamble defines the system's initial data, with the following values:
	
	\begin{center}
		\begin{tabular}{|| c | c | l||}
			\hline
			Variable	&	Value	&	Meaning	\\
			\hline\hline
			$D$	&	$10$	&	Initial puncture separation	\\
			\hline
			$q$	&	$36.0 / 29.0$	&	Mass ratio $(q \geq 1)$	\\
			\hline
			M	&	$1.0$		&	Total mass	\\
			\hline
			$\vec \chi_p$	&	$[0,0,0.31]$	&	Primary BH spin	\\
			\hline
			$\vec \chi_m$	&	$[0,0,-0.46]$		&	Secondary BH spin\\
			\hline
			$p_r$	&	$-0.00084541527$		&	Radial linear momentum\\
			\hline
			$p_{\phi}$	&	$0.09530152297$		&	Azimuthal linear momentum\\
			\hline
			
		\end{tabular}
	\end{center}
	
	It is important to note, that all calculations within this parameter file - and by extension - within most of the toolkit, are carried out in Geometrized units \cite{Natural}, whereby the convention is taken that $G = c = 1$, therefore implying that all distances and times are given in units of mass. Furthermore, the convention is typically taken that the total mass ($M = m_p + m_m$) is equal to one. The mass ratio ($q = \frac{m_p}{m_m}$) is one of great importance, influencing the behaviour of the inspiral and subsequent merger in a profound manner. Here it is defined to be greater than one. $\chi_p \text{ and } \chi_m$ are the three-dimensional dimensionless intrinsic spins of the two black holes respectively, defined by $\vec \chi = \vec J / M^2 $ \cite{spin}, where $\vec J$ is the black hole's angular momentum and it is known that $0 \leq |\vec \chi| \leq 1$.
	
	Following this, the remainder of the preamble sets up various necessary values such as the estimated duration of the merger, time steps of each iteration as well as setting up the grid structure used during the simulation itself.
	
	The parameter file then lists all thorns required for the simulation, activating and subsequently specifying operational parameters. These are described within the thorn guide, provided at \url{https://einsteintoolkit.org/documentation/ThornGuide.php}. Here, a few important thorns will be discussed in order to paint a picture of how the simulation will proceed.
	
	Several of the thorns, including !Coordinates!, !Interpolate2! and !GlobalDerivative! are parts of the !llama! multi-block infrastructure \cite{Pollney:2009yz}. The purpose of these thorns is to work in unison to allow the use of multiple curvilinear coordinate grid patches for !Cactus!, using grids provided by !Carpet!. Here, !Thornburg04! is a system consisting of 7 patches. These include a central cubic Cartesian grid capable of adaptive mesh refinement, surrounded by 6 "inflated cube" spherical grids, together covering a spherical shell. !WaveExtractCPM! is also particularly important, providing a method to extract gauge-invariant perturbations of the Schwarzschild metric away from the binary itself. These data then allow for the study of gravitational waves being emitted throughout the simulation's evolution. 
	
	As mentioned above, !Carpet! is a mesh refinement driver \cite{Schnetter:2003rb} which is able to run with multiple grid patches. It runs by actively altering the refinement of areas of the grid, providing greater sensitivity and accuracy to sensitive or turbulent regions - as compared to other, less interesting ones.
	
	The !AHFinderDirect! thorn quickly and accurately locates apparent horizons of black holes based on provided initial guesses \cite{Thornburg:2003sf}. The local apparent horizon is the surface boundary which separates outward directed light rays travelling outwards from outward directed light rays travelling inwards. It therefore acts as the local notion of a black hole's boundary, giving a good description of the puncture's position.
	
	!TwoPunctures! is a thorn which provides initial data for a system with two black holes \cite{Ansorg:2004ds} (two punctures) located along the x-axis, with opposite linear momenta along the y-axis. It does so by utilising a spectral method in solving the necessary Hamiltonian constraint equations as described by \cite{Ansorg:2004ds}.
	
	The !QuasiLocalMeasures! thorn allows for the real-time calculation of quasi local measures such as masses, linear and angular momenta on a closed two-dimensional surface \cite{Dreyer:2002mx}. These data are then outputted to allow the simulation's progress to be tracked.
	Following this, the !ML_BSSN! thorn is used to evolve the simulation, by iteratively solving a BSSN-type formulation of the Einstein Field Equations \cite{Brown:2008sb}, using a collection of fourth-order accurate integration methods.
	
	Unsurprisingly, the Output section gives details of which data is produced, at which intervals. Furthermore, it is specified which format is used to store the data. Specifically, !IOASCII! and !IOHDF5! are used, with the rationale described in the next section.
	
	\section{Using Toolkit Data}\label{data}
	
	Simulations such as the BBH Merger above produce a huge amount of data, with the full simulation producing several terabytes of data \cite{wardell_barry_2016_155394}. However, even with lightweight data (comprising a small number of iterations) plenty of information can be gleamed, as shown in the figures below.
	
	In general, all data produced in simulations using the Einstein Toolkit are saved in one of two formats: ASCII (!.asc!) and HDF5 (!.h5!). Both formats have their advantages and disadvantages and are therefore used, though for different data types.
	
	Data stored as !.asc! files are immediately accessible on all computing platforms, appearing as a typical !.txt! file. These can immediately be parsed by python, using a tool such as the \textbf{pandas} package \cite{mckinney-proc-scipy-2010}. Although this is doable, it is recommended to use a short script to convert the data into a !.csv! (comma separated values) file, for ease of use:
	\begin{lstlisting}[language=bash]
	#!/bin/sh
	cat puncturetracker-pt_loc..asc | grep "^[^#;]" > foo1.txt
	sed -e 's/ /\t/g' foo1.txt > foo2.txt
	sed -e 's/\t/,/g' foo2.txt > puncturetracker-pt_loc.csv
	
	echo done
	\end{lstlisting}
	
	Here, data produced by the !PunctureTracker! thorn is converted into a !.csv! format, while simultaneously formating the data to be comma-delimited and removing all commented lines. This modified file can then be read in using the following lines in python:
	\begin{lstlisting}[language=python]
	import pandas as pd
	data = pd.read_csv("puncturetracker-pt_loc.csv", names=headers)
	time = pd.DataFrame.to_numpy(data['time'])
	\end{lstlisting}
	where !headers! is a 1x52 array containing the data column headers as described in the original !puncturetracker-pt_loc..asc! file. The final line then extracts the 'time' column, converting it into a numpy array - which can then easily be used for any further analysis.
	
	On the other hand, data saved as a !.h5! file are stored in a hierarchical format, which allows for the storage of huge amounts of data with a high degree of compression. Furthermore, this format allows for very fast Input/Ouput (I/O) as compared to ASCII. However, in order to visually inspect and edit data stored in a !.h5! format, it is necessary to use an interface tool such as HDFView \cite{HDFView}. 
	
	In order to parse !.h5! data using python, the \textbf{h5py} package \cite{h5py} is required. The following code will then allow you to read in data:
	\begin{lstlisting}[language=python]
	import h5py
	
	path = 'GW150914_28/output-0000/GW150914_28/mp_psi4.h5'
	filename = l2_m2_r500.00
	hf = h5py.File(path, 'r')
	data = hf[filename]
	\end{lstlisting}
	
	This code will open the example simulation's Weyl Scalar ($\Psi_4$) \cite{weyl} information and extract data for the (2,2) spherical harmonic mode. The resultant array will have three columns - giving the Real and Imaginary components as a function of time.
	
	Following this, it is clearly possible to plot and visualize the simulation data. Here, several examples are shown using the above BBH simulation data \cite{wardell_barry_2016_155394}. Data was parsed using code similar to that which was shown above, and all post-processing and plotting were done using the freely accessible packages \textbf{numpy}, \textbf{scipy} and \textbf{matplotlib.pyplot} \cite{Hunter:2007, numpy, scipy}.

	\begin{center}
		\begin{figure}[h!]
			\begin{subfigure}{0.35\textwidth}
				\hspace{2.5cm}
				\includegraphics[width = \linewidth]{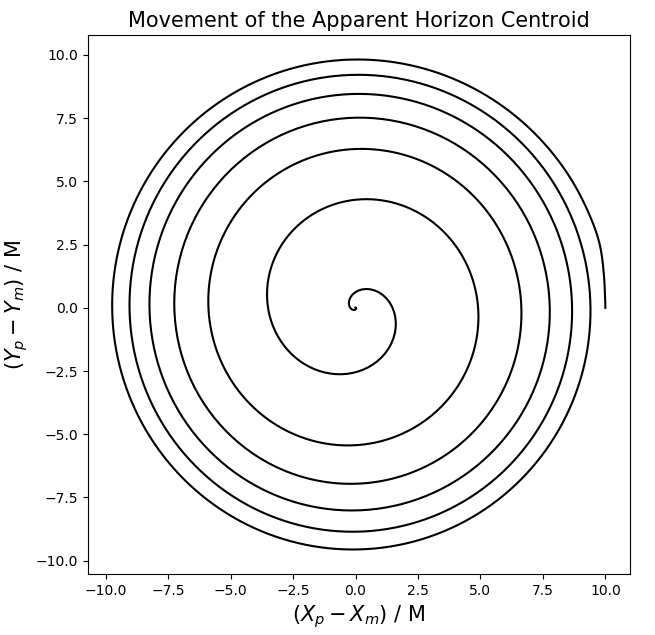}
			\end{subfigure}
			\begin{subfigure}{0.35\textwidth}
				\hspace{2.5cm}
				\includegraphics[width = \linewidth]{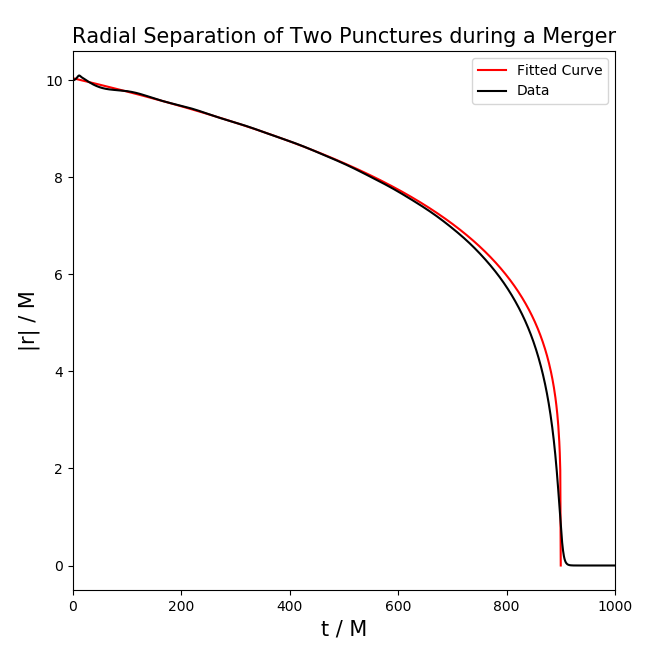}
				
			\end{subfigure}
			
			\begin{subfigure}{0.7\textwidth}	
				\hspace{2.2cm}
				\includegraphics[width = \linewidth]{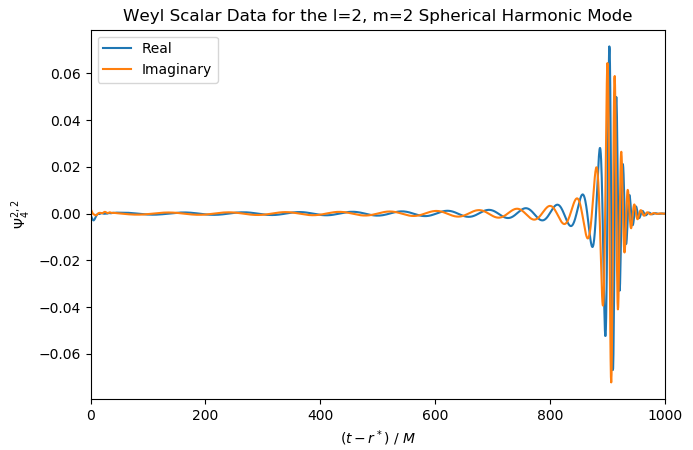}
			\end{subfigure}
			\caption{\protect\label{CombinationFig}Several plots of data produced in a simulation of the binary black hole merger GW150914 using the Einstein Toolkit. \textit{Top left:} Movement of the black holes' centroid throughout the simulation is shown. The inspiral stage is clearly seen, with the rate of convergence increasing as the two black holes approach one another. \textit{Top right:} Here the precise radial separation between the two black holes is shown as a function of time. This data was then fitted, with the result having the expected form. \textit{Bottom:} Raw Weyl Scalar data for the dominant (2,2) spherical harmonic mode, as measured at a distance of $r = 500M$. This shows the expected increase in both frequency and amplitude culminating in the merger itself.}
		\end{figure}
	\end{center}

	Figure \ref{CombinationFig} shows several plots of the raw simulation data. These include the motion of the system's centroid, the change in radial separation as a function of time as well as raw data from the Weyl Scalar for the dominant (2,2) spherical harmonic mode. The first of these clearly demonstrates the expected quasicircular nature of the inspiral, with an innermost stable circular orbit of $r_{\text{ISCO}} \approx 6.6M$. The second shows a rapid increase in the rate of radial approach as a function of separation. This data was fitted to an equation of the form: $r(t) = r_0 (1 - t/t_{\text{coalesce}})^b$ with the linear regression algorithm yielding $b = 0.2361 \pm 0.0003$. For binary systems it is possible to show that the averaged power emitted by gravitational waves is given by \cite{celoria2018lecture}
	\begin{equation}
	\frac{dE}{dt} = - \frac{32}{5} \frac{\mu^2 M^3}{r^5} F(e) = \frac{32}{5}\frac{M_1^2 M_2^2 (M_1 + M_2)}{r^5} F(e),
	\end{equation}
	where the factor
	\begin{equation}
	F(e) = (1 - e^2)^{-7/2} \bigg{(} 1 + \frac{73}{24}e^2 + \frac{37}{96}e^4 \bigg{)}.
	\end{equation}
	Here, taking the Newtonian approximation that $E = \frac{- M_1 M_2}{2r}$ and that $e \approx 0$ gives
	\begin{equation}
	\frac{dr}{dt} = - \frac{64}{5} \frac{M_1 M_2 (M_1 + M_2)}{r^3},
	\end{equation}
	which can easily be integrated to give $r(t) = r_0 (1 - t/t_{\text{coalesce}})^{1/4}$. Therefore, the simulation's orbital separation decreases slightly faster than this approximation, as would be expected given the relativistic nature of the system.
	
	Following this, the bottom plot of figure \ref{CombinationFig} shows the variation in the Weyl Scalar of the dominant (2,2) spherical harmonic mode, as measured at a distance of $r = 500M$. This demonstrates the expected sharp rises in both frequency and amplitude, as well as the complete 'chirp' \cite{LIGOObs} itself, as the two black holes merge.
	
	In order to calculate the measurable strain ($h$) - as would be measured by a ground-based detector such as LIGO -  we use the following formalism \cite{Integral}
	\begin{equation}
	h = h_+ - i h_{\text{$\times$}} = \int_{- \infty}^t dt' \int_{- \infty}^{t'} dt'' \Psi_4,
	\end{equation}
	where $h_+$ and $h_{\times}$ are the plus and cross polarizations respectively, as they appear in the transverse-traceless gauge \cite{Polarization}. In order to carry out this integration, and to account for a linear drift we use the procedure outlined in \cite{Integral}. Doing so yields figure \ref{22modeFig} below.
	
	Furthermore, for figure \ref{22modeFig} we have accounted for the propagation time taken for the signal to reach the detector at $r = 500M$. This was done by approximating the propagation time by $t \approx r^*$, where $r^*$ is the Regge-Wheeler "tortoise coordinate" \cite{tortoise}
	\begin{equation}
	r^* = r + 2M \ln \bigg{|} \frac{r}{2 M} - 1\bigg{|}.
	\end{equation}
	 
	\begin{center}
		\begin{figure}[h!]
			\includegraphics[width=0.9\linewidth]{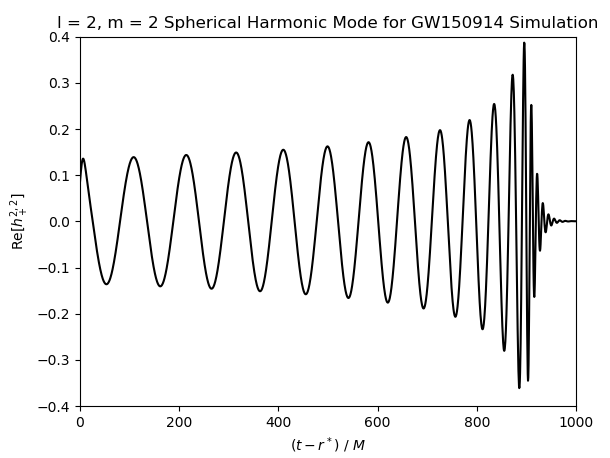}
			\caption{\protect\label{22modeFig}Plot of the dominant $h_+^{2,2}$ strain, as would be measured by a gravitational wave observatory such as LIGO. Key features include the constant-amplitude and frequency waves produced by the initial inspiral, the rapidly increasing frequency and amplitude of the merger itself, as well as the equally rapid ringdown, showing the aftermath of the binary black hole merger. The entire signal lasted 0.2s, with a peak gravitational wave frequency of 250Hz.}
		\end{figure}
	\end{center}
	Figure \ref{22modeFig} shows the build-up and chirp in strain of the $h_+^{2,2}$ spherical harmonic mode, as produced by the GW150914 inspiral, merger and ringdown. These stages are characterized by the initial approximately constant amplitude and frequency, followed by a rapid increase in both frequency and amplitude, followed by an equally rapid drop in both during the ringdown phase. Accounting for units, the entire signal would have lasted 0.2 seconds, with a peak amplitude at a frequency of 150Hz - as detected by LIGO \cite{LIGOObs}.  
	
	Following the merger, the resultant black hole was inferred to have a total mass of $62^{+ 4}_{- 4} M_{\odot}$, implying that $3.0^{+0.5}_{-0.5} M_{\odot}c^2$ was radiated as gravitational waves \cite{LIGODATA}, such as those shown in figure \ref{22modeFig}. The successful detection of GW150914 was a landmark event in gravitational wave astronomy. It provided the first evidence of the existence of binary stellar-mass black hole systems, as well as being the first direct detection of gravitational waves.	

	\section{Conclusion} 
	
	To conclude, this document has described the process of installing, building and subsequently using the Einstein Toolkit, from the perspective of a student. The aim being to dispel some of the 'mystery', and to show students that the world of numerical relativity is not only fascinating, but entirely feasible to start exploring, without the need for a computing cluster, or even heavy technical skills.
	
	Furthermore, the document has explored several examples of numerical simulations done using the toolkit, with the focus being on a model of a binary black hole merger. This was done using system properties inferred by LIGO of the GW150914 event. Several plots were then produced - with the process described and explained.
	
	The feasibility of utilising AWS computing infrastructure in order to carry out computational physics has also been considered. Though an expensive tool, AWS is able to provide an extraordinary amount of computing power - which a student would have otherwise found difficult to obtain. This being the case, more work must to be done to benchmark AWS capabilities, in order to ascertain the optimal setup for uses of this kind.
	
	As a result, it has been reiterated that the Einstein Toolkit is a powerful tool, allowing for both high-level simulations as well as experimentation with numerical relativity 'at home'. It has also been shown that it is not only entirely feasible for students to use the Toolkit, but equally that it is an incredible opportunity for those students to learn and gain experience of using numerical relativity.
	
	\bibliography{ETTutorial}
	\bibliographystyle{ieeetr}
	
\end{document}